\documentclass[prb,twocolumn,showpacs,showkeys]{revtex4}%
\usepackage{graphicx}
\usepackage{amsmath}
\usepackage{amsfonts}
\usepackage{amssymb}
\usepackage{graphicx}
\usepackage{epsfig}%
\providecommand{\U}[1]{\protect\rule{.1in}{.1in}}
\providecommand{\U}[1]{\protect\rule{.1in}{.1in}}
\begin{document}
\author{Alejandro Cabo-Bizet$^{1}$ and Alejandro Cabo Montes de Oca$^{2,3}$ }
\title{Particle dynamics in a relativistic invariant stochastic medium. }
\affiliation{$^{1}$ Facultad de F\'{\i}sica, Universidad de La
Habana, Colina Universitaria, La Habana, Cuba. }
\affiliation{$^{2}$ Grupo de F\'{\i}sica Te\'{o}rica, Instituto de
Cibern\'{e}tica, Matem\'atica y F\'{\i}sica, La Habana, Cuba}
\affiliation{$^{3}$Abdus Salam International Center for
Theoretical Physics, Strada Costiera 11, Miramare, Trieste,
Italy.}

\begin{abstract}
\noindent The dynamics of particles moving in a medium defined by
its relativistically invariant stochastic properties is
investigated. For this aim, the force exerted on the particles by
the medium is defined by a stationary random variable as a
function of the proper time of the particles. The equations of
motion for a single one-dimensional particle are obtained and
numerically solved. A conservation law for the drift momentum of
the particle during its random motion is shown. Moreover, the
conservation of the mean value of the total linear momentum for
two particles repelling each other according with the Coulomb
interaction is also following. Therefore, the results indicate the
realization of a kind of stochastic Noether theorem in the system
under study.  Possible applications to the stochastic
representation of Quantum Mechanics are advanced.

 \pacs{02.50.Ey,03.65.Ud,05.30.Ch,05.40.-a}
\keywords{Brownian motion, stochastic quantization, EPR,
measurement}
\end{abstract}
\maketitle

\section{ Introduction}

Stochastic processes made their appearance in research in Physics
long time ago and their theory has played an important role in the
description of systems which do not behave in a deterministic
manner \cite{vanKampen,wio,Zaslavski}. In particular, the study of
the dynamics of particles lying inside material media has been the
object of \ high interest. A classical example is the study of the
Brownian motion \cite{vanKampen}. A large deal of those
investigations had a non-relativistic character and the random
interactions with the background medium were considered as being
dependent of the state of motion of the particle, that is, lacking
invariance under the changes of the reference system
\cite{vanKampen,wio,Zaslavski}. Another large class of \ studies
in this field  had been directed to show the equivalence with
random processes of the solutions of quantum relativistic or
non-relativistic equations, like the Klein-Gordon, Dirac and
Schrodinger ones. \cite{
nelson,kershaw,vigier,bohm,deAngelis,guerra,boyer,delaPena,
torres}. Two basic additional subjects  in connection with
stochastic processes in Quantum Theory are:  The  attempts to
derive the 'collapse' of the wave function during measurements
from the existence of random perturbations in Quantum Mechanics
(QM) \cite{pearle,ghirardi,adler}, and  the study of the
decoherence processes and their role in spontaneous  transitions
from pure to mixed states \cite{zurek}.

The  main objective of the present work is to investigate some
consequences on the motion of a particle determined by the action
exerted over it by a medium which random properties are defined in
a relativistically invariant form. The basic motivation is simple:
It is recognized that the Copenhagen interpretation of Quantum
Mechanics (QM), is the most successful and dominant from all the
existing ones.\ However, it  is also accepted that its approach to
measurements constitute one its more puzzling aspects, which up to
now is widely debated in the literature\cite{
nelson,kershaw,vigier,bohm,deAngelis,guerra,boyer,delaPena,
torres,pearle,ghirardi,adler,zurek}. Let us suppose for a moment,
that in opposition to the Copenhagen interpretation and in
accordance with Einstein expectations, the  particles in Nature
are in fact localized at definite points of the space at any
moment. Then, the only way we can imagine for the quantum
mechanical properties of the motion to emerge from a model, is
that the action of the vacuum on the particles have a stochastic
character. But, the relativistic invariance of the vacuum, leads
to expect that the acceleration felt by the particle in its proper
frame should be a stationary random variable as a function of the
proper time. This circumstance motivates the study started here
about the motion of particles inside a random media showing the
above mentioned property. For the sake of simplicity the one
dimensional motion is considered. \ It is not attempted to show
the equivalence of the dynamics in the medium with the one
predicted by the quantum mechanical equations. The purpose in this
first step, being redundant, is to study general properties of the
motion of one and two particles assuming two main conditions: a)
They have a definite localization in the space at any moment and
b) The forces acting on the particles have the above mentioned
random properties which are independent the observer's inertial
reference frame.

The work will proceed as follows. Firstly, the equations of motion
of the particles under the action of the medium are formulated.
For this purpose the properties which ensure the relativistic
invariance of the\ motion under the action of the medium are
stated by specifying the form of the random forces. Further, the
equations of motion of a single particle are written and solved \
and a statistical analysis of \ the random properties is done. \ A
main conclusion following is the existence of a \ conservation law
for a mean drift momentum and kinetic energy of a 'free' particle
propagating in the medium. \ It \ indicates the validity of a kind
of stochastic Noether theorem which links the relativist
invariance of the stochastic motion with the conservation \ of \
the mean 4-momentum of the particle.

\ Further, \ the conservation law \ is studied \ for the mean of the addition
of two four momenta associated to the scattering of two identical particles,
which repel each other through an instantaneous \ Coulomb interaction. It is
concluded that the random action of the medium does not modify the usual
conservation law, valid for the impact in the absence of external forces. \

A review of the results and future extensions of the work are
presented in a conclusion section. Some possibilities to extend
the study are advanced. In general terms, our view is that the
form of the analysis have the chance of being useful in the search
for consistent hidden variables models. The study of these
possibilities will be  considered elsewhere.

\section{Equation of motion}

In this section \ we will obtain and solve \ the Newton equation
of motion for a particle on which a random force $F_{p}(\tau)$ is
\ acting. \ A\ one dimensional system will be considered to make
the discussion as simple as possible. The force will be defined as
a vector in the proper reference frame of the particle and will
depend on the proper time $\tau.$ \ That means, in each instant we
will consider an inertial reference frame moving relative to the
observer's fixed frame (Lab frame) with the velocity of the
particle $\nu$ and having the origin of coordinates coinciding
with it. \ In this system of reference,
after a time increment $d\tau$, it is possible to write%
\begin{equation}
F_{p}(\tau)\text{ }d\tau=m_{0}\text{ }d\nu^{\prime}, \label{Fp}
\end{equation}
where\ \ $m_{0}$ is the proper mass of the particle.

The particle reaches  a small velocity $d\nu^{\prime}$ relative to
this system and
a new velocity respect to the Lab frame $\nu+d\nu$, given by the equation%
\begin{align}
\nu+d\nu &  =\frac{\nu+d\,\nu^{\prime}}{1+\frac{\nu\ \ d\nu^{\prime}}{c^{2}}%
},\nonumber\\
&  \cong(\nu+d\,\nu^{\prime})\times(1-\frac{\nu\text{ }d\,\nu^{\prime}}{c^{2}%
}),\\
&  \cong\nu+(1-\frac{\nu^{2}}{c^{2}})\text{ }d\,\nu^{\prime},
\label{teoria_nueva velocidad en tierra}%
\end{align}
where $c$ is the velocity of light. Thus, the variation of speed in the Lab frame $d\nu$ is%
\begin{equation}
d\,\nu=(1-\frac{\nu^{2}}{c^{2}})\text{ }d\,\nu^{\prime}. \label{dif_vel}%
\end{equation}

From expressions (\ref{Fp}) and (\ref{dif_vel}) the required differential
equation for the motion is obtained :%
\begin{eqnarray}
F_{p}(\tau)&=&\frac{m_{0}}{(1-\frac{\nu^{2}}{c^{2}})}\frac{d\,\nu}{d\,\tau},
\label{Dif_Mov} \\
\nu &=& \frac{d\text{ }x(t)}{d\text{
}t}=(1-\frac{\nu^{2}}{c^{2}})\frac{d\text{
}x}{d\text{ }\tau}.%
\end{eqnarray}

It is useful to state the relation between the strength of the
force in the Lab system and its proper frame counterpart, which
is:
\begin{align}
F_{p}(\tau)  &  =\sqrt{1-\frac{\nu^{2}}{c^{2}}}F_{L}(\tau
)\label{trans_fuerza}.
\end{align}
\ \ However, since the relativistic invariance condition will be
imposed on $F_{p}(\tau)$ this will be the type of force  mostly
considered in what follows. Integrating the equation \ref{Dif_Mov}
in the proper time it follows
\begin{align*}
\int F_{p}(\tau)\text{ }d\tau+\hat{C}  &  =m_{0}\int\frac{1}{(1-\frac{\nu^{2}%
}{c^{2}})}\text{ }d\nu,\\
&  =\frac{m_{0}\text{\thinspace}c}{2}\text{ }\ln(\frac{1+\frac{\nu}{c}%
}{1-\frac{\nu}{c}}),
\end{align*}
which determines the velocity the Lab frame $\nu(\tau)$\ \ as a
function of the proper time $\tau$, only through the dependence of
$\tau,$ of the integral of the random force $F_{p}(\tau).$ The
explicit form of $v$ \ becomes
\begin{equation}
\nu(\tau)=c\cdot\tanh[\frac{1}{m_{0}c}\cdot(\int F_{p}(\tau)\cdot d\tau
+\hat{C})], \label{velocidad_en_tiempo_propio}%
\end{equation}
where $\hat{C}$ is an arbitrary constant.

\section{The random force}

As it was mentioned in the Introduction, the medium under study
will be defined in the proper frame as randomly acting \ over the
particle being at rest in it. \ That is, its action in this
reference system will be given \ by \ a stochastic process showing
no preferential spatial direction  and assumed to be produced by
an external relativistic system which dynamics is unaffected by
the presence of the particle. Its is  also natural to impose the
coincidence of the distribution function of the forces of the
medium for a large sampling interval of proper time $T$ \ and the
one obtained fixing the proper time $\tau,$\ produced by an
ensemble of a large number of samples of the forces taken during
long time intervals $T$. \
\begin{figure}[h]
\epsfig{figure=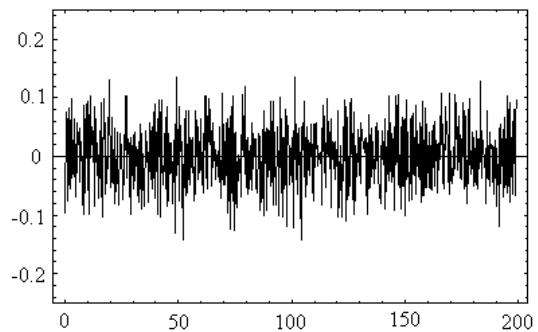,width=8cm} \vspace{-0cm}\caption{ The
figure shows a {\it realization} of the force field corresponding
 to a  spectrum of $N=200$ frequencies. The horizontal coordinate is
the time in seconds
and the vertical one is  the force in Newtons. The amplitude was
fixed $f_0=1 \;\; (N_t) \; \; (N_t=Newton)$. }%
\label{graficofuerzas1}%
\end{figure}

These conditions, can be assured by a random force $F_{p}(\tau)$ \
being stationary, ergodic and symmetrical distributed about the
zero value of the force. A numerical realization of a band limited
white noise distribution obeying these properties is implemented
in reference \cite{Kafadar} and will be employed here. Concretely,
the expression for the stochastic force given in the proper
reference frame as a function of the proper time will be taken in
the form
\begin{align}
f_{N}(\tau,\varphi_{i}) &  =\frac{f_{0}}{N}\sum_{i=1}^{N}\cos(w_{i}(N)\text{
}\tau+\varphi_{i}),\label{fuerza_estocastica}\\
w_{i}(N) &  =\frac{8\pi i}{N },\;\;\;  i=1,...N ,
\end{align}
where the $N$ \ phases $\varphi_{i},$ $i=1,...N,$ are randomly
chosen with a uniform distribution between $0$ and $2\pi$. The
integer number $N$ is the number of frequency components of the
numerical band limited white noise distribution. The  bandwidth
will be chosen as a fixed one and equal to \ $\Delta w=8\pi$.
Clearly, the exact randomness for an arbitrarily large time
interval only will attained in the infinite limit of $N$. The
parameter $f_{0}$ controls the amplitude of the force values. \
Note that the absence of a zero frequency component is implied by
the condition of the process not showing a preferential direction
in space.
\begin{figure}[h]
\epsfig{figure=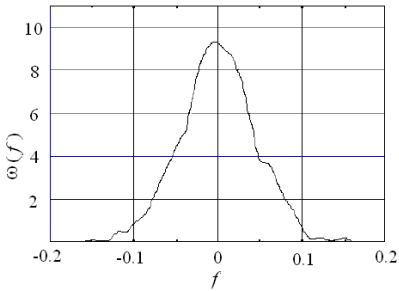,width=6cm} \vspace{-0cm}\caption{ The
distribution function of the ensemble of forces for the set of
frequencies $\ w_{i}=\frac{8\pi} {250}i$ \
$(s^{-1})$,$\;i=1,...250$ and $f_{0}=$ $1$ $\ (N_{t})$.}%
\label{distribucion fuerza}%
\end{figure}

Figure \ref{graficofuerzas1} shows the force distribution $f_{N}$
for a $realization$ of the ensemble of forces \ for $\ N=200.$ A
$realization$ here is called the time evolution of the force, for
the set of randomly fixed phases $\varphi_{i}$ at the start. The
picture qualitatively shows the stationary character of the random
force.
 Figure \ref{distribucion fuerza} depicts an
interpolation curve of the data for the distribution function
$w(f)$ \ corresponding to an ensemble of \ random
\textit{realizations} of the force. The white noise frequency
spectrum was defined by $N=250$ frequency components $\ w_{i}=\frac{8\pi}%
{250}i$ \ $(s^{-1})$,$\;i=1,...250$.

The force amplitude fixed was as $f_{0}=$ $1$ $\ (N_t).$\ \ \ The
picture corresponds to a large sampling time $T$  ( it is
sufficient to be at most equal to  the period of the smallest
frequency of the spectrum). Notice the even character of the
distribution and its rapid decay to zero. \ Of course, this occurs
inside the interval defined by $\ f_{1,2}=\pm1$ $N_{t}.$. \ This
result is \ natural due to the fact that the force is normalized
and its absolute value can not exceed $\ f_{0}$.

\section{A particle in the medium}

 In this Section the existence of a conservation law for the mean
momentum (to be also called from now on the drift momentum) will
be shown for a particle moving in the before defined random
medium. \ To this purpose \ let us employ the solution \
(\ref{velocidad_en_tiempo_propio}) which was found \ in the
previous section \ We will combine \ this expression with the
results obtained from the definition of the random force ensemble
\ in equation (\ref{fuerza_estocastica}), to explicitly determine
$v=v(\tau).$ \ The resulting relation links the velocity of the
particle in the Lab \ frame with the proper time measured by a
clock fixed to it. \ Once having $v(\tau)$ we will comment about
its stationary random behavior. The existence of a non vanishing
conserved mean value of the velocity $<v>$ will be shown. \
Starting form the method \ defined above, \ the velocity
distribution function will be determined for the frequency
spectrum defined before and \ few representative values of the
arbitrary constant $\hat{C}$. After that, the position of the
particle $x(t)$ as a function of the time $t$ \ \ measured at the
Lab frame will be evaluated.
\begin{figure}[h]
\epsfig{figure=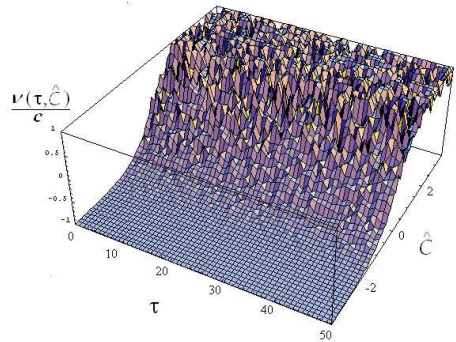,width=6cm} \vspace{-0cm}\caption{ The
velocity of the particle in the laboratory frame against the
proper time and the arbitrary constant $\hat{C}$. The parameters
for the random force were $\frac{f_{0} }{8\pi m_{0}c}=0.1\,
($s$^{-1}$), $\ w_{i}=\frac{8\pi\,\,i}{N}$  (s$^{-1})$,
$\ i=1,...,250$.}%
\label{graficovel1}%
\end{figure}

Taking $F_{p}(\tau)$ to be given by the white noise force \ $f_{N}(\tau)$ \ we
have :%
\begin{equation}
F_{p}(\tau)=f_{N}(\tau)=\frac{f_{0}}{N}\sum_{i=1}^{N}\cos(\frac{8\pi i\text{
}}{N}\tau+\varphi_{i}). \label{fuerza2}%
\end{equation}
Then, integrating with respect to $\tau$ \ produces
\begin{align}
I_{F_{p}}(\tau)  &  =\int_{0}^{\tau} F_{p}(\tau)\text{
}d\tau=\frac{f_{0}}{N}\int_{0}^{\tau}
[\sum_{i=1}^{N}\cos(\frac{8\pi i\text{ }}{N}\tau+\varphi_{i})]\text{ }%
d\tau,\nonumber\\
&  =\frac{f_{0}}{8\pi}[\sum_{i=1}^{N}\frac{1}{i}\sin(\frac{8\pi i\text{ }}%
{N}\tau+\varphi_{i})]. \label{integral_fuerza}%
\end{align}

Substituting (\ref{integral_fuerza}) into (\ref{velocidad_en_tiempo_propio}), gives%
\begin{align}
\nu(\tau)  &  =c\text{ }\tanh[\frac{1}{m_{0}c}(I_{F_{p}}(\tau)+\hat{C})],\\
&  =c\text{ }\tanh[\frac{1}{m_{0}c}(\frac{f_{0}}{8\pi}[\sum_{i=1}^{N}\frac
{1}{i}\sin(\frac{8\pi i}{N}\tau+\varphi_{i})]+\hat{C})]. \label{v(tao)}%
\end{align}

\begin{figure}[h]
\epsfig{figure=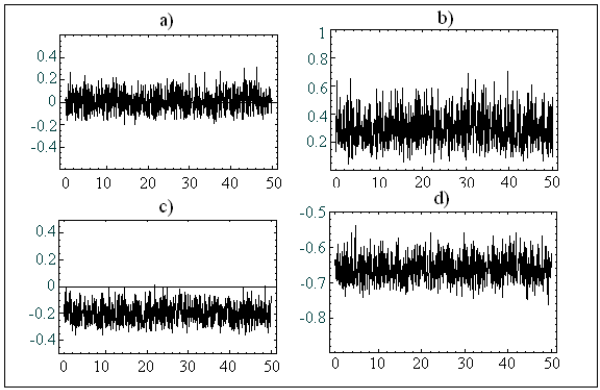,width=8cm} \vspace{-0cm}\caption{ The
velocities of the particles (divided by $c$) in the laboratory
system vs. the proper time for four specific values of $\hat{C}$:
\ a) $\hat{C}$=0 \, b) $\hat{C}$=0.3 \ c) $\hat{C}$=-0.2 and d)
$\hat{C}$=-0.8. Note the non vanishing value of the mean velocity
for $\hat{C}$ different from zero.}%
\label{graficovel2}%
\end{figure}

It can be seen that the summation within the argument of the
hyperbolic tangent is symmetrically distributed \ around its zero
value. Since the $\tanh(x)$ is an anti-symmetric function, it
follows that when $\hat{C}\neq0$ there will be a nonvanishing mean
velocity of the particle in the medium. As the mean value of the
kinetic energy is also conserved, it can be said that the mean
4-momentum of the particle is conserved. Moreover, the
relativistic invariance of the system also implies that the mean
4-momentum of the solutions for any two values of $\hat{C}$ should
be linked by certain Lorentz transformation. Thus, the particle
trajectories for the various values of $\hat{C}$  are simply a
fixed trajectory (in the stochastic sense) after being Lorentz
transformed into certain moving frame.

The picture in figure \ref{graficovel1} shows \ the behavior of
$\nu
=\nu(\tau,\hat{C})$ for the following values of the parameters $\frac{f_{0}%
}{8\pi m_{0}c}=0.1\,$s$^{-1}$, $\ w_{i}=\frac{8\pi\,\,i}{N}$
s$^{-1}$, $\ i=1,...,250$. \ \ \ Note that for fixed values of
$\hat{C}$ (that means, for certain initial conditions) the
velocity rapidly oscillates around \ a value being close to the
quantity $\tanh[\hat{C}]$. It illustrates  the mentioned
conservation of the mean drift velocity for the particle in spite
of the random oscillations of the medium. \ Figure
\ref{graficovel2} shows the same dependence for specific values of
$\hat{C}$.

\subsection{Distributions}

\begin{figure}
 \epsfig{figure=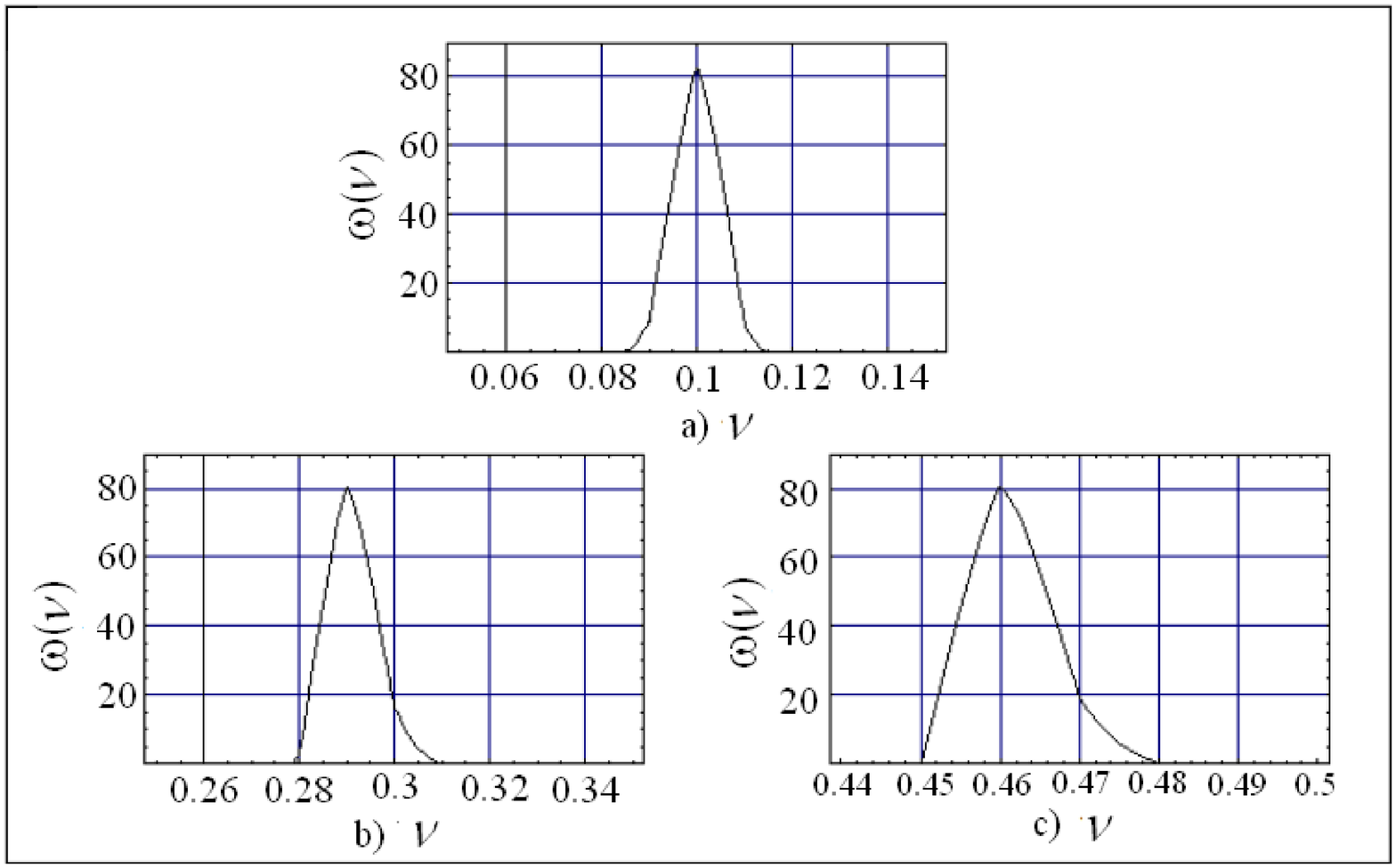,width=8cm}
\vspace{-0cm}\caption{ The distribution functions:  a) w(v)  for
$\hat{C}$=0.1, \ b) w(v) for $\hat{C}$=0.3  and c) w(v), for
$\hat{C}$=0.5. Note the distortion of the symmetry around the
center, when $\hat{C}$ grows.}   %
\label{distribuciones}%
\end{figure}

Let us consider now the numerical evaluation of the distribution
function
 $w(v)$ (for measuring a given value of the  velocity $v$) for few
 representative values of $\hat{C}$.
\ This function is found on the basis of the ergodic property  of
the system, that is, by performing \ a large number of evaluations
of the velocity's expression with time running from zero to a
"sampling" value $T$, for afterwards compute \ the number of
times, for which  $v$ \ takes values in a small neighborhood of a
given value. Further,\ after interpolating the results, the
distribution functions are obtained after dividing by the fixed
size of the mentioned neighborhood. \ In figure
(\ref{distribuciones}) the distribution  $w(v)$ is plotted for a
few values of $\hat{C}$. \ \

These pictures correspond to the frequency spectrum
$w_{i}=\frac{8\pi}{250}i,\;\;i=1...250,\;$ but with the amplitude
fixed by $\frac{f_{0}\text{ }125}{4\pi}=1.$ The almost
independence of the form of these curves on the size of a large
"sampling" time $\ T$ employed (whenever $N$  is sufficiently
large) indicates the approximate validity of the ergodic property
of the white noise implementation employed.

Note the rapid decay away from the mean value and the deformation
of the symmetry around it, \ of the distribution $w(v)$ for
different values of $\hat{C}$. \ These distributions allow to \
calculate the mean velocities in each of the cases. \ This method
will be employed in the next sections for finding the mean values
as integrals over the domain of the quantity,  of its value times
the distribution function.

\begin{figure}[h]
\epsfig{figure=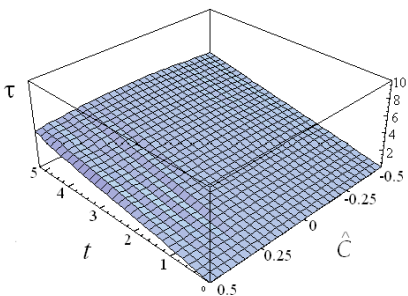,width=6cm} \vspace{-0cm} \caption{
Dependence on $t$ of  the proper time in the laboratory
$\tau=\tau(t,\hat{C})$,  with $\hat{C}$ running along a third
axis. Note that, since the dependence, by definition, always
should be monotonous, the random oscillation are not apparent.
They are however  present in the local slope of the curves.  }%
\label{tiempos}%
\end{figure}

\begin{figure}[h]
\epsfig{figure=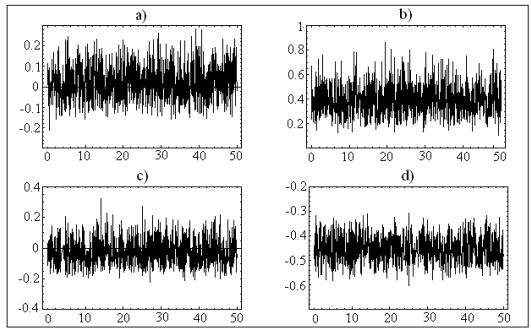,width=7cm} \vspace{-0cm}\caption{
 The velocities in the laboratory frame (divided by c),   but now plotted as functions of the time $t$ in
  this same frame of reference, for the values of $\hat{C}$: \ a) $\hat{C}$ = 0.01, b)
$\hat{C}$=0.4, c) $\hat{C}$=-0.03, d)
$\hat{C}$=-0.5. }%
\label{velocidad de la particula libre}%
\end{figure}

\subsection{\ Particle velocities and trajectories}

\ It can be noted that the explicit solution for the velocity
obtained in (\ref{velocidad_en_tiempo_propio})\ corresponds to
this velocity in terms of the proper time $\tau$. However, in
order to integrate the velocity to find the random particle
trajectories as functions of the time in the laboratory frame, it
becomes necessary to know the functional relationship
$\tau(t,\hat{C})$ for the considered trajectory. Then, let us
consider now \ \ the numerical determination of this relation. The
parameters\ of the random force will be $\frac{f_{0}}{8\pi
m_{0}c}=$\ $0.1\,($s$^{-1})$\ , $w_{i}=\frac{8\pi\,\,i}{N}$
(s$^{-1})$, $i=1,2,...,250.$ \ \ Using the solution
$\nu=\nu(\tau,\hat{C})$ it follows
\begin{align}
t(\tau,\hat{C})  &  =\int\limits_{0}^{\tau}\frac{d\tau^{\prime}}%
{(1-\frac{v(\tau\prime,\hat{C})^{2}}{c^{2}})^{\frac{1}{2}}},\label{t(tao)}\\
\;t(0)  &  =0,
\end{align}
and employing it, the values of $t(\tau,\hat{C})$ were numerically
found. Afterwards, finding the inverse mapping $\
\tau=\tau(t,\hat{C})$, this function is  depicted in  figure
\ref{tiempos} with $\hat{C}$ as a parameter.
Finding the composed function for various values of $\hat{C}$ ,
the observer's time dependence of the particle velocity $\
\nu(t,\hat{C})$ \ follows. The results are illustrated in figure \
 \ref{velocidad de la particula libre}, for the chosen values of $\hat{C}$.

\begin{figure}[h]
\epsfig{figure=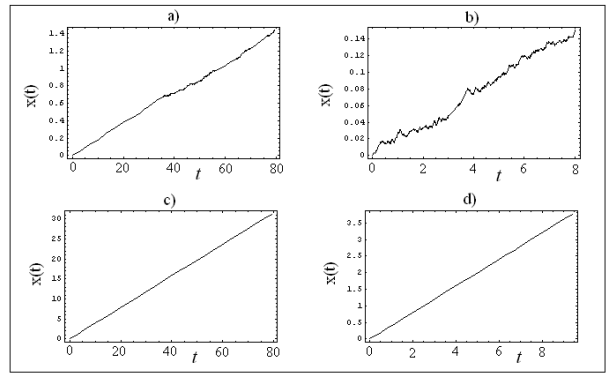,width=8cm} \vspace{-0cm}\caption{The
particle coordinates divided by $c$: $\frac{x}{c}=\frac{1}{c}x(t)$
as functions of the time in the laboratory frame. In a) and b)
$\hat{C}$ = 0.01 but the time scale is different.  For c) and d)
$\hat{C}$=0.4 and again the two scales are different. Note that
for $\hat{C}$ small the randomness seems to be greater than for
larger
$\hat{C}$. }%
\label{posicion de la particula}%
\end{figure}

Further, after integrating the calculated velocities with respect
to $\ t$ $\ $(starting at $t=0$ \ and imposing $x(0)=0$) \ the
values of the positions $x(t)$ respect to the laboratory reference
frame are obtained. \ \ The trajectories of the particles are
shown in  figure \ref{posicion de la particula}, for the same set
of values of $\hat{C}.$

Note the randomness of the motion in the case of small absolute
values of $\hat{C}$ \ which is not the case for the larger ones. \
This property can be understood analyzing the pictures in figure
(\ref{velocidad de la particula libre}). In the cases in which the
velocity constantly changes from positive to negative values and
vice versa, the randomness is more evident. Thus, the aleatory
effect is more visible in one case than in the other, because the
curve changes from a monotonous to non monotonous behavior. \ But,
after taking into account the relativistic invariance of the
model, as noticed before, it can be seen that the curves for \
different values of $\hat{C}$ should transform any one into
another by \ a Lorentz transformation (exactly, when the same
\textit{realization} of the force is employed, or statistically,
if another \textit{realization} is used). Therefore, the apparent
increased randomness for low $\hat{C}$ is only a visual effect
enforced by the change of sign of the velocity in frames in which
its mean value is sufficiently low. \

\section{Two particles in the medium}

\ Let us consider in this final section the conservation of the
mean total momentum of \ a 'closed' system of two particles which
travel in the medium by also interacting between them. A repulsive
interaction between the two identical particles $1$ and $2$ will
introduced. \ \ \ The forces between the particles will be defined
in the laboratory frame for studying the impact between them
there. They will be chosen as satisfying the Law of Action and
Reaction \ and having the Coulomb form. If, for example, the
forces have an electromagnetic origin, the retardation effects
will be disregarded in order to assure that the field \ will not
deliver momentum to the set of two particles. This approximation
is appropriate for low particle velocities in its random motion in
the Lab frame.

The system of differential equations will be written and
numerically solved. \ The results for the position and velocity of
both particles $x_{1}(t)$, $v_{1}(t)$, $x_{2}(t)$, $v_{2}(t)$,
will show how after the impact, any of them deliver to the other
its mean drift velocity and its type of randomness. As it was
mentioned above, the expression for
the repulsive force in the Lab frame will have the Coulomb form%
\begin{equation}
\overrightarrow{F}_{rep_{1}}(x_{1},x_{2})=-\overrightarrow{F}_{rep_{2}}%
(x_{1},x_{2})=\frac{\alpha}{|x_{1}-x_{2}|^{3}}(\overrightarrow{x}%
_{1}-\overrightarrow{x}_{2}). \label{repulsiva}%
\end{equation}

Consider now the relativistic Newton equations \ for both
particles in their respective proper frames and also the two
transformations between the common laboratory time $t$ \ and the
two different proper times $\tau_{1}$ and $\tau_{2}.$ \ These
relations may be written as
\begin{widetext}
\begin{align}
\frac{m_{0}}{(1-\frac{^{x_{1}^{\prime}(t)^{2}}}{c^{2}})^{\frac{3}{2}}}%
\frac{d^{2}x_{1}}{dt^{2}}  &  =F_{p_{1}}(\tau_{1})+(1-\frac{^{x_{1}^{\prime
}(t)^{2}}}{c^{2}})^{\frac{1}{2}}F_{rep_{1}}(x_{1},x_{2}),\;\label{sistema1}\\
\frac{m_{0}}{(1-\frac{x_{2}^{\prime}(t)^{2}}{c^{2}})^{\frac{3}{2}}}\frac
{d^{2}x_{2}}{dt^{2}}  &  =F_{p2}(\tau_{2})+(1-\frac{x_{2}^{\prime}(t)^{2}%
}{c^{2}})^{\frac{1}{2}}F_{rep_{2}}(x_{1},x_{2}),\;\label{sistema2}\\
\frac{d\tau_{1}}{dt}  &  =(1-\frac{x_{1}^{\prime}(t)^{2}}{c^{2}})^{\frac{1}%
{2}},\;\;\;\tau_{1}(t_{0})=\tau_{1_{0}},\;x_{1}(t_{0})=x_{1_{0}}%
,\;x_{1}^{\prime}(t_{0})=v_{1_{0}},\label{sistema3}\\
\frac{d\tau_{2}}{dt}  &  =(1-\frac{x_{2}^{\prime}(t)^{2}}{c^{2}})^{\frac{1}%
{2}},\;\;\;\tau_{2}(t_{0})=\tau_{2_{0}},\;x_{2}(t_{0})=x_{2_{0}}%
,\;x_{2}^{\prime}(t_{0})=v_{2_{0}}. \label{sistema4}%
\end{align}
\end{widetext}
In order to be consistent with the non retarded approximation for
the Coulomb repulsion, as mentioned above, initial velocities for
the particles \ being relatively small with respect to $c$ were
taken. \ This assumption makes the considered approximation to be
satisfied if the intensity of the repulsive force is weak so that
the velocities after the impact will also be small.
\begin{figure}[h]
\epsfig{figure=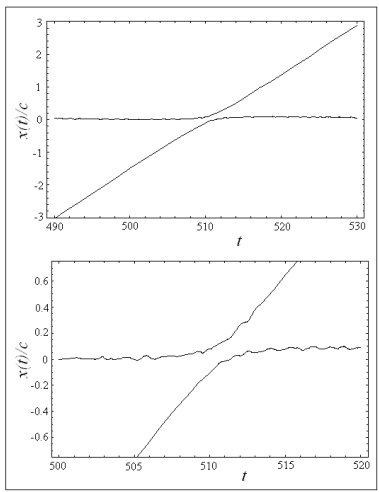,width=5cm} \vspace{0.25cm} \caption{
 The trajectories of the particles $\ x_{1}(t)$ and $x_{2}(t)$ during the impact for
 the repulsion parameter  $\alpha=0.01\ (N\times m^{2})$ and initial conditions
 $x_{1}(500)=x_{2}(500)=0$, $v_{1}(500)=0.015$, $v_{2}(500)=0.15$,
$\tau_{1}(500)=0$, $\tau_{2}(500)=0$.\ The curve  at the bottom is
the same that the one at the top, but at a lower scale in the vertical coordinate. }%
\label{choque1}%
\end{figure}
\begin{figure}[h]
\epsfig{figure=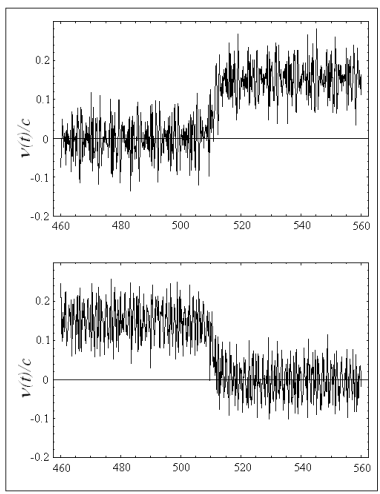,width=5cm} \vspace{-0cm} \caption{
The time dependence for  the velocities (divided by $c$)  of both
particles: a) $v_{1}(t)/c$ and  b)$\ v_{2}(t)/c$ in the considered
shock. Note, the exchange of
their mean velocity values.   }%
\label{choque2}%
\end{figure}
Employing the white noise parameters $w_{i}=\frac{8\pi}{250}i$, \
$i=1,...250$ \ and $\;\frac{f_{0}\text{ }125}{4\pi}=1$, we
obtained the numerical solutions shown in figure \ref{choque1} for
various initial conditions. \ \ The conservation law of the total
mean momentum of the two particles can be noticed. Figure
\ref{choque2} \ shows the time variation of the velocities of the
particles. The process of exchange of their drift momenta is
clearly illustrated and therefore the corresponding conservation
of the total momentum. This outcome follows in spite of the \
strong oscillations \ produced by the action of the medium, in
opposition to what happens within a standard material media. In
these systems the stochastic action normally tends to stop the
motion of the particle making the drift velocity to vanish,
assumed the absence of accelerating external fields. The
calculated values for the mean velocities before and after the
shock are
\begin{eqnarray}
<v_{1_{0}}>&=&0, \; \; \; \;\;  \;  \; <v_{2_{0}}>=0.152, \\
<v_{1_{f}}>&=&0.152, \; \; <v_{2_{f}}>=0  ,
\end{eqnarray}
 which numerically confirms the
conservation of the total momentum in the shock of two particles
forming a 'closed' system immersed and moving in a relativistic
random medium.

\section{Conclusions}

The one-dimensional equation of motion for a particle moving in a
medium having relativistic invariant stochastic properties is
formulated and numerically solved. \ The \ velocity of the
particles in the medium is a function of the proper time only
through the integral of the force in the proper reference frame.
This relation directly shows \ the existence of a stochastic
conservation law:
\textit{ A free particle in the defined random
medium conserves its mean momentum and kinetic energy \ along its
motion.}\

The problem of two particles moving in the medium is also
investigated  after considering a momentum conserving Coulomb
repulsion between them. The evaluated solutions generalize the
conservation law: \textit{The sum of the two drift momenta of two
particles moving in the medium without any other external action
conserves after a shock.  }

An interesting outcome is that for two different shocks with
identical initial conditions for both particles when they are far
apart, the drift velocities before the impact are not uniquely
determined by the initial conditions. These velocities also depend
on the array of phases utilized for the \textit{realization }\ of
the force.\textit{ }Therefore, it follows that the \ results of
the impact will show a  dependence on the specific
\textit{realization} of the random force.  \ This circumstance
implies that the probability distribution associated to an
ensemble of particles all situated at the beginning at the same
point and with a fixed value of the velocity, \ will not define
only one direction of the trajectory for large times. \ The
probability distributions of such an ensemble, on the other hand,
should evolve in space and time \ in a relativistically invariant
manner, as the Lorentz invariance of the system indicates. The
same conclusion can be traced for any other sort of fixed boundary
conditions. Therefore, the set of possible initial conditions for
the particle (considering also that they can be placed in the
medium at different initial spacial and temporal points for the
construction of the ensemble) should be expected to be equivalent
to the set of all possible space time evolving ensembles that can
be observed. \ The above remarks suggest some possible extensions
of the present work, which are described below:
\begin{itemize}
\item The  indicated dependence of the ending results of the
shocks, not only of the initial conditions, but also of the
concrete \textit{realization} of the random force, \ led us to
think in extending the results to the 2D and 3D cases. The idea is
to study the spatial and temporal behavior of the ensemble of
outcomes o a series of shocks "prepared " with fixed initial mean
velocities for both particles,  and to compare the results  with
the corresponding predictions of the quantum scattering.

\item Another task of interest \ which is suggested by this study
is to investigate the existence of preferential bounded states in
the case that the Coulomb interaction is considered as attractive.
$\ $A particular simpler situation could be to assume one of the
particles as very massive, that is,  merely acting  as an
attracting center. \ In both cases the study could consist in \
determining the statistical properties of the stationary
trajectories given the initial condition. \ \ \
\end{itemize}

\bigskip

In conclusion, we would like to underline \ that it seems not
without sense that the realization of the above proposed \ studies
could be of use in the justification or search for \ hidden
variable approaches to Quantum Mechanics (QM). As it can be seen
from the discussion, the resulting models could not show the
limitations of the Bohm approach (like the absence of predictions
for the \ "guided" motions of the particles for real wave
functions, for example). Moreover, the analysis seems of interest
in seeking for a theory not requiring to fix a particular random
process to each stationary state, but one in which all the
statistical properties can emerge from a general framework. In the
imagined outcome, the particle could propagate having a
probability for transit or stay into each one of \ a set of
stationary random motions which could be associated (but now
within the general context) to the particular eigenfunctions of
the system. As for the allowance of the necessary \ \
generalizations needed to make contact with reality, it can
guessed that the possible generalization of the statistical
Noether properties (detected in the simple model considered here)
could lead to the conservation of the \ mean values for the \
angular momentum and other internal quantities, in analogy with
the case of the linear momentum. We think these possibilities \
deserve  further examination that will be published elsewhere.

\end{document}